%% file: main.tex
\documentclass[pdflatex,sn-mathphys-num]{sn-jnl}

\usepackage{graphicx}%
\usepackage{multirow}%
\usepackage{amsmath,amssymb,amsfonts}%
\usepackage{bm}
\usepackage{amsthm}%
\usepackage{mathrsfs}%
\usepackage[title]{appendix}%
\usepackage{xcolor}%
\usepackage{textcomp}%
\usepackage{manyfoot}%
\usepackage{booktabs}%
\usepackage{algorithm}%
\usepackage{algorithmicx}%
\usepackage{algpseudocode}%
\usepackage{listings}%
\usepackage{comment}
\usepackage{makecell}
\usepackage{fancybox}
\usepackage{tabu, booktabs}
\usepackage{todonotes}
\usepackage{siunitx}
\usepackage{tabularx}
\usepackage{multirow}       
\usepackage{array}          
\usepackage{subfigure}
\usepackage{hyperref}
\usepackage{tikz}
\usetikzlibrary{shapes.geometric, arrows.meta, positioning, patterns,calc}

\usetikzlibrary{arrows.meta, calc, decorations.pathreplacing,
                ext.paths.ortho, positioning, quotes}
\makeatletter
\tikzset{phantom node/.code=\tikz@addoption{\expandafter\let\csname pgf@sh@boxes@\tikz@shape\endcsname\pgfutil@empty}}
\makeatother
\tikzset{
  shadowed node xshift/.initial=1.5ex, shadowed node yshift/.initial=1ex, shadowed node list/.initial={2, 1},
  pics/shadowed node/.default=\pgfkeysvalueof{/tikz/shadowed node list},
  shadowed node/.pic={
    \foreach[expand list] \elem in {#1}
      \scoped[transparency group, shadowed node calculation={\elem}]
        \node[style/.expand once=\tikzpictextoptions, phantom node,
              xshift={\elem*\pgfkeysvalueof{/tikz/shadowed node xshift}},
              yshift={\elem*\pgfkeysvalueof{/tikz/shadowed node yshift}}] (-\elem) {\tikzpictext};
    \node[alias=-0, style/.expand once=\tikzpictextoptions] () {\tikzpictext};},
  set shadowed node calculation parameter/.style={shadowed node calculation/.style={opacity={(#1-##1+1)/(#1+1)}}},
  set shadowed node calculation parameter=2,
  overshoot line to/.style={to path={($(\tikztostart)!-(#1)!(\tikztotarget)$)--($(\tikztotarget)!-(#1)!(\tikztostart)$)\tikztonodes}},
  edges have transparency group/.style={execute at begin to={\scope[transparency group,#1]}, execute at end to=\endscope}}

\newcommand{\jingquan}[1]{\todo[color=purple!30, inline]{\textbf{Jingquan:} #1}}

\definecolor{hanblue}{rgb}{0.27, 0.42, 0.81}

\definecolor{codegray}{gray}{0.9}
\definecolor{codered}{rgb}{0.6,0,0}

\lstdefinestyle{mystyle}{
    backgroundcolor=\color{codegray},
    commentstyle=\color{codered},
    basicstyle=\ttfamily\footnotesize,
    breaklines=true,
    postbreak=\mbox{\textcolor{red}{$\hookrightarrow$}\space},
    showstringspaces=false
}

\theoremstyle{thmstyleone}%
\theoremstyle{thmstyletwo}%

\theoremstyle{thmstylethree}%

\begin{document}

\title[Article Title]{ChronoLLM: A Framework for Customizing Large Language Model for Digital Twins generalization based on PyChrono}


\author[1]{\fnm{Jingquan} \sur{Wang}}\email{jwang2373@wisc.edu}

\author[1]{\fnm{Harry} \sur{Zhang}}\email{hzhang699@wisc.edu}

\author[1]{\fnm{Khailanii} \sur{Slaton}}\email{kslaton@wisc.edu}

\author[1]{\fnm{Shu} \sur{Wang}}\email{swang597@wisc.edu}

\author[1]{\fnm{Radu} \sur{Serban}}\email{serban@wisc.edu}

\author[1]{\fnm{Jinlong} \sur{Wu}}\email{jinlong.wu@wisc.edu}

\author*[1]{\fnm{Dan} \sur{Negrut}}\email{negrut@wisc.edu}

\affil*[1]{\orgname{UW-Madison}, \orgaddress{\street{1513 University
Avenue}, \city{Madison}, \postcode{53706}, \state{WI}, \country{USA}}}


\abstract{Recently, the integration of advanced simulation technologies with artificial intelligence (AI) is revolutionizing science and engineering research. "ChronoLlama" introduces a novel framework that customizes the open-source LLMs, specifically for code generation, paired with PyChrono for multi-physics simulations. This integration aims to automate and improve the creation of simulation scripts, thus enhancing model accuracy and efficiency. This combination harnesses the speed of AI-driven code generation with the reliability of physics-based simulations, providing a powerful tool for researchers and engineers. Empirical results indicate substantial enhancements in simulation setup speed, accuracy of the generated codes, and overall computational efficiency. ChronoLlama not only expedites the development and testing of multibody systems but also spearheads a scalable, AI-enhanced approach to managing intricate mechanical simulations. This pioneering integration of cutting-edge AI with traditional simulation platforms represents a significant leap forward in automating and optimizing design processes in engineering applications.}




\maketitle
\section{Introduction}\label{sec:Introduction}

\input{sections/Introduction.tex}

\section{Methodology}\label{sec:Methodology}
\input{sections/Methodology}

\section{Numerical Test}\label{sec:Numericaltest}
\input{sections/Numericaltest}

\section{Conclusion}\label{sec:Conclusion}
\input{sections/Conclusion}

\section{Limitations}
\input{sections/Limitation}

\section{Future Work}
\input{sections/Future_work}

\bmhead{Supplementary information}

All codes, data, and models used in this study will be open-sourced on GitHub. The link will be provided in the final version of the paper.


\section*{Declarations}
All codes, data, and models used in this study will be open-sourced on GitHub. The link will be provided in the final version of the paper.

\begin{appendices}
\input{sections/Appendix.tex}

\section{Prompts and Data Synthesis Pipeline}\label{secA1}




\end{appendices}


\bibliography{BibFiles/refsChronoSpecific,BibFiles/refsSBELspecific,BibFiles/refsML-AI}

\end{document}

%% file: sections/Introduction.tex
\subsection{Project Chrono and Impact of Chrono}\label{sec:chrono}

Project Chrono \cite{tasora2016chrono} is an open-source, physics-based simulation framework that supports the modeling, simulation, and analysis of complex systems. It is designed for high-performance, high-fidelity simulations and is widely used in research and industry. PyChrono \cite{pychrono2022} is the Python wrapper for Project Chrono, providing a user-friendly interface to the core functionalities of Project Chrono. It allows users to leverage the power of Project Chrono using Python, making it accessible to a broader range of users who prefer scripting in Python over C++.

Project Chrono encompasses a wide range of features, and PyChrono inherits a subset of these capabilities:

\begin{enumerate}
    \item Chrono::Engine: Provides core functionality for multibody dynamics and nonlinear finite element analysis, with robust treatment of friction and contact using both the penalty method and the Lagrange-multiplier method.
    \item Chrono::Cascade: Enables interoperability with CAD tools, allowing the import of mechanical systems defined in SolidWorks into Chrono.
    \item Chrono::Vehicle: Includes a comprehensive library of wheeled and tracked vehicles, facilitating high-fidelity vehicle dynamics simulations, engine modeling, terrain-tire interaction, and deformable terrain simulations. It focuses on off-road and unstructured scenarios involving deformable terrains and movable obstacles.
    \item Chrono::ROS: A ROS bridge that enables Chrono to interact with ROS-based robots, such as Phoenix. 
    \item Chrono::Sensor: Provides sensor modeling and simulations.
    \item Chrono::Parsers: A tool to import external models and to interact with other languages like URDF, OpenSim, and Adams.
\end{enumerate}

Chrono has been utilized by NASA in conjunction with the 2024 VIPER mission, which aims to search for frozen water on the Moon. It has also been adopted by the Department of Defense High-Performance Computing Modernization Program (HPCMP) for the simulation of wheeled and tracked vehicles in the CREATE-GV project \cite{skorupaGVSETS2017}. Additionally, Chrono has been tested in NATO benchmarking exercises for off-road vehicle mobility analysis \cite{NATObenchmark2018}. Other notable users include the Jet Propulsion Lab, U.S. Army, Argonne National Lab, National Higher French Institute of Aeronautics and Space, Riken (Japan), over 100 universities, and companies such as Caterpillar, Oshkosh Corporation, Mitsubishi Heavy Industries, British Aerospace Engineering, and more. Applications of Chrono span extraterrestrial exploration \cite{toso2015esa,ferrari2017n,jplREACH2020,regolithSamplingChrono2023}, machine learning in robotics \cite{cookNeuralNets2017}, image processing \cite{mccormacChrono2017,xu2018ComputerVisionChrono}, autonomous vehicles \cite{batteryAutonomousVehiclesChrono2016,goodin2017unmanned,haraus2017performance,trajectoryPlanningEpureanu2023,tractionControlChronoHungary2024,roboticFishChrono2024}, tracked-vehicle design \cite{jonakTrackedVehicle2018}, fluid-solid interaction \cite{dualSPHCouplingChrono2016}, bridge suspension \cite{wangCableChrono2019}, hardware-in-the-loop simulations \cite{HILchrono2019}, wind turbine dynamics \cite{perezChrono2018,marten2019benchmark}, hydrodynamics \cite{ogden2023hydrochrono} and oil industry applications \cite{oilIndustryChrono2020}.

As of May 2024, Project Chrono has been developed for 25 years and iterated to its 9.0.0 version, boasts over 650 forum members \cite{projectChronoForum}, the PyChrono module has been downloaded more than 11,000 times \cite{pyChronoCondaWebSite}, and the Chrono Docker image has been pulled over 1,800 times \cite{projectChronoDockerImage}. Additionally, Chrono has been starred with more than 2.1k times and forked more than 400 times on GitHub \cite{projectChronoGithub}, resulting in numerous derivative projects.

\subsection{LLMs and Domain-Specific LLMs}\label{sec:domainLLMs}

Recent advancements in artificial intelligence, particularly in large language models (LLMs), have led to significant breakthroughs in natural language processing. The scalability of these models, as highlighted by scaling laws \cite{kaplan2020scalinglaw1,hoffmann2022scalinglaw2}, demonstrates that as models increase in size, they exhibit emergent abilities that were not evident at smaller scales. These emergent abilities \cite{wei2022emergent1,schaeffer2024emergent2} include enhanced comprehension, reasoning, and language generation, paving the way for LLMs to extend their utility beyond simple text tasks to more complex applications across various domains.

In the realm of science and engineering, LLMs are significantly transforming how professionals approach problem-solving and design. Advanced closed-source LLMs, such as the GPT family \cite{OPENAI2024gpt4,NIPS_2020} by OpenAI, the Gemini family \cite{team2023gemini} by Google, and the Claude family \cite{Claudemodelcard} by Anthropic, have demonstrated substantial progress in handling complex tasks like code generation and system simulation, which are crucial in engineering applications. The usage of these closed-source LLMs is typically through websites and API calls, which are charged services. While these LLMs, trained on trillions of tokens from publicly available data, perform exceptionally well on general tasks, they have notable limitations.

One significant limitation is the lack of domain-specific knowledge. The extensive general training data of closed-source LLMs often lacks the depth required for specialized domains. Moreover, due to their large size, frequent retraining to incorporate recent developments is impractical. This lack of exposure to up-to-date information can lead to inaccurate responses, particularly in domain-specific information processing where LLMs may not fully understand new terminology.

To build an LLM that can help with domain-specific problems, it is essential to teach LLMs with specialized knowledge, enabling them to adapt and comprehend new information accurately. This targeted approach helps mitigate the limitations of closed-source LLMs, ensuring that they remain relevant and precise in specialized applications.

\begin{table}[h]
\centering
\begin{tabular}{|l|p{3.5cm}|p{3.5cm}|p{3.5cm}|}
\hline
& \textbf{Prompt Engineering} & \textbf{Prompt Learning} & \textbf{Fine-Tuning} \\ \hline
\textbf{Tech} & 
\begin{itemize}
\item Few-shot learning \cite{wang2020fslearning} 
\item Chain-of-thought reasoning \cite{wei2022NIPS_ChainOfThought}
\item Scratchpad reasoning \cite{google2021scrathpad}
\end{itemize} & 
\begin{itemize}
\item Prompt tuning \cite{gu2021ppt}
\item P-tuning \cite{liu2021ptun}
\item Prefix tuning \cite{li2021prefix}
\end{itemize} & 
\begin{itemize}
\item Adapters \cite{hu2023adapter}
\item LoRA \cite{hu2022lora}
\item RLHF \cite{bai2022rlhf}
\end{itemize} 
\\ \hline
\textbf{Pros} & 
\begin{itemize}
\item Lowest investment 
\item Least expertise
\item Retains old skills
\end{itemize} & 
\begin{itemize}
\item Lower investment 
\item Better performance
\item Retains old skills
\end{itemize} & 
\begin{itemize}
\item Best performance
\end{itemize}  \\ \hline
\textbf{Cons} & 
\begin{itemize}
\item Worst performance
\item Slow down inference
\item Limited addition of
skills or data

\end{itemize} & 
\begin{itemize}
\item Slow down inference
\item Limited addition of
skills or data
\end{itemize} & 
\begin{itemize}
\item Medium investment 
\item Longer training 
\item More expertise 
\item Forget old skills
\end{itemize}  \\ \hline
\end{tabular}
\caption{Comparison of Methodologies to customized LLMs from pre-trained LLMs}
\label{tab:fine_tune_methods}
\end{table}

Generally speaking, there are four different ways to build an LLM to work for domain-specific tasks.

\begin{enumerate}
\item Training from Scratch: This approach involves collecting extensive domain-specific data and training a new LLM tailored to address specific problems. For instance, in the field of high-performance computing, models such as OMPGPT-0.78B \cite{chen2024ompgpt} and MonoCoder-0.89B \cite{kadosh2023domain} have claimed superior performance compared to GPT-3.5 in parallelizing C++ code using OpenMP. The primary advantage of this method is the potential for achieving the highest performance when sufficient data, computational resources, and expertise are available. However, the disadvantages are significant, including the high cost of data collection, training, and computational resources. Training even a small-scale 7-8B LLM like LLaMA3-8B \cite{llama3modelcard} can cost millions of dollars. If the domain-specific LLM is too small, its performance on specific tasks will be highly constrained.
\item Prompt Engineering or In-Context Learning: This approach uses domain-specific examples (few-shot learning) and well-crafted prompts to enhance the reasoning capabilities of LLMs (e.g., chain-of-thought \cite{wei2022NIPS_ChainOfThought} and ScratchPad \cite{google2021scrathpad}). By leveraging the strong abilities of closed-source LLMs, we can customize them for domain-specific tasks. For example, in the multibody dynamics field, researchers have used GPT-4 to generate model codes \cite{gerstmayr2024multibodyllm} for the Exudyn library \cite{gerstmayr2023exudyn}. Other notable examples include MyCrunchGPT \cite{kumar2023mycrunchgpt} for generating Python codes for physics-informed neural networks (PINN) \cite{RAISSI2019PINN} and GeoGPT \cite{KIM2024geogpt} for generating Matlab code in geotechnical engineering. The advantages of this approach are the minimal expertise and computational resources required, as no model parameters are altered, and the strong performance of GPT-4 on general tasks. However, the limitations include potential security issues due to the need to upload information to the internet, limited customization capabilities, and high API costs and inference latency.

\item Prompt Learning: This method goes beyond prompt engineering by introducing additional trainable parameters to generate better prompts, such as prefix-tuning \cite{li2021prefix}, P-tuning \cite{liu2021ptun}, and prompt tuning \cite{gu2021ppt}. For example, prompt learning has been used to customize LLMs for clinical diagnosis \cite{taylor2023clinical}. The pros of this approach are its low training cost and the fact that the LLM parameters remain unchanged, often yielding better results than simple prompt engineering. The cons include the limited ability to improve performance without altering all model parameters and potential inference slowdowns.

\item Fine-Tuning and Parameter-Efficient Fine-Tuning: Besides the powerful closed-source LLMs, excellent open-source LLMs with available weights, such as the LLaMA family \cite{meta2023llama,touvron2023llama2,roziere2023codellama,llama3modelcard} by Meta, Mistral family \cite{jiang2023mistral} by Mistral AI, Gemma family \cite{google2024gemma,codegemma_2024} by Google, and Phi family \cite{abdin2024phi3} by Microsoft, can be fine-tuned. Fine-tuning involves starting from pre-trained LLM weights and continuing training with domain-specific data. This method is considered one of the best for customizing LLMs for specific problems, as demonstrated by various studies. For example, OceanGPT \cite{bi2023oceangpt} is trained based on LLaMA3-8B for ocean-related problems, and Hippocrates \cite{acikgoz2024healthcarellm} fine-tuned Mistral-7B and LLaMA2-7B, showing superior performance compared to larger models up to 70B. AnomalyGPT \cite{gu2023anomalyagpt} fine-tuned large vision-language models for anomaly detection in manufacturing processes. C4QGPT, supporting query generation for quantum dynamics \cite{aragones2024c4qgpt}, is fine-tuned based on BERT \cite{google2018bert}, and MPIrigen \cite{schneider2024mpirigen} fine-tunes MonoCoder to parallelize C++ code using MPI. The advantages of fine-tuning include achieving the best results based on pre-trained LLMs, the ability to incorporate extensive data, and not slowing down inference. However, fine-tuning requires some level of training, necessitates expertise, and may lead to forgetting old skills or even catastrophic forgetting (CF). Research \cite{zhao2024loraland} shows that parameter-efficient fine-tuning (PEFT) with open-source 7-8B LLMs can outperform GPT-4 on domain-specific tasks by an average of 10 points. There are also solid researches \cite{liu2022finetunevsicl,mosbach2023finetunevsicl2} prove that the performance improvement by fine-tuning is much higher than prompt engineering. We will talk more about fine-tuning and parameter-efficient fine-tuning in the next section.
\end{enumerate}

\begin{table}[ht]
\centering
\begin{tabular}{|c|c|c|c|c|}
\hline
\textbf{LLMs} & \textbf{Size} & \textbf{Base Model} & \textbf{Method} & \textbf{Domain} \\
\hline
OPENMPGPT\cite{chen2024ompgpt} & 0.78B & N/A & Trained from Scratch & High-Performance Computing \\
\hline
MonoCoder\cite{kadosh2023domain} & 0.89B & N/A & Trained from Scratch & High-Performance Computing \\
\hline
MyCrunchGPT\cite{kumar2023mycrunchgpt} & N/A & GPT-4 & Prompt Engineering & PINN \\
\hline
Exudyn with GPT-4\cite{gerstmayr2024multibodyllm} & N/A & GPT-4 & Prompt Engineering & Multibody Dynamics \\
\hline
GeoGPT\cite{KIM2024geogpt} & N/A & GPT-4 & Prompt Engineering & Geotechnical Engineering \\
\hline
Clinical Prompt learning\cite{taylor2023clinical} & N/A & N/A & Prompt Learning & Health Care\\
\hline
C4QGPT\cite{aragones2024c4qgpt} & 2x0.11B & Bert\cite{google2018bert} & Fine-tuning & Quantum Dynamics \\

\hline
MPIRIGEN\cite{schneider2024mpirigen} & 0.89B & MonoCoder & Fine-tuning & High-Performance Computing \\
\hline
OceanGPT\cite{bi2023oceangpt} & 8B & LLaMA3-8B & Fine-tuning & Ocean Engineering \\
\hline
Hippocrates\cite{acikgoz2024healthcarellm} & 7B & LLaMA2-7B & Fine-tuning & Health Care \\
\hline
\end{tabular}
\caption{Overview of LLMs for domain-specific problems.}
\label{tab:engineering_llms}
\end{table}

For convenience, We conclude the advantages and disadvantages of different customization methods with a pre-trained LLM in table \ref{tab:fine_tune_methods} and the mentioned LLMs for domain-specific problems in table \ref{tab:engineering_llms}.

\subsection{PyChrono Challenges and current LLMs}

\begin{figure}[htbp]
    \centering
    \includegraphics[width=12cm]{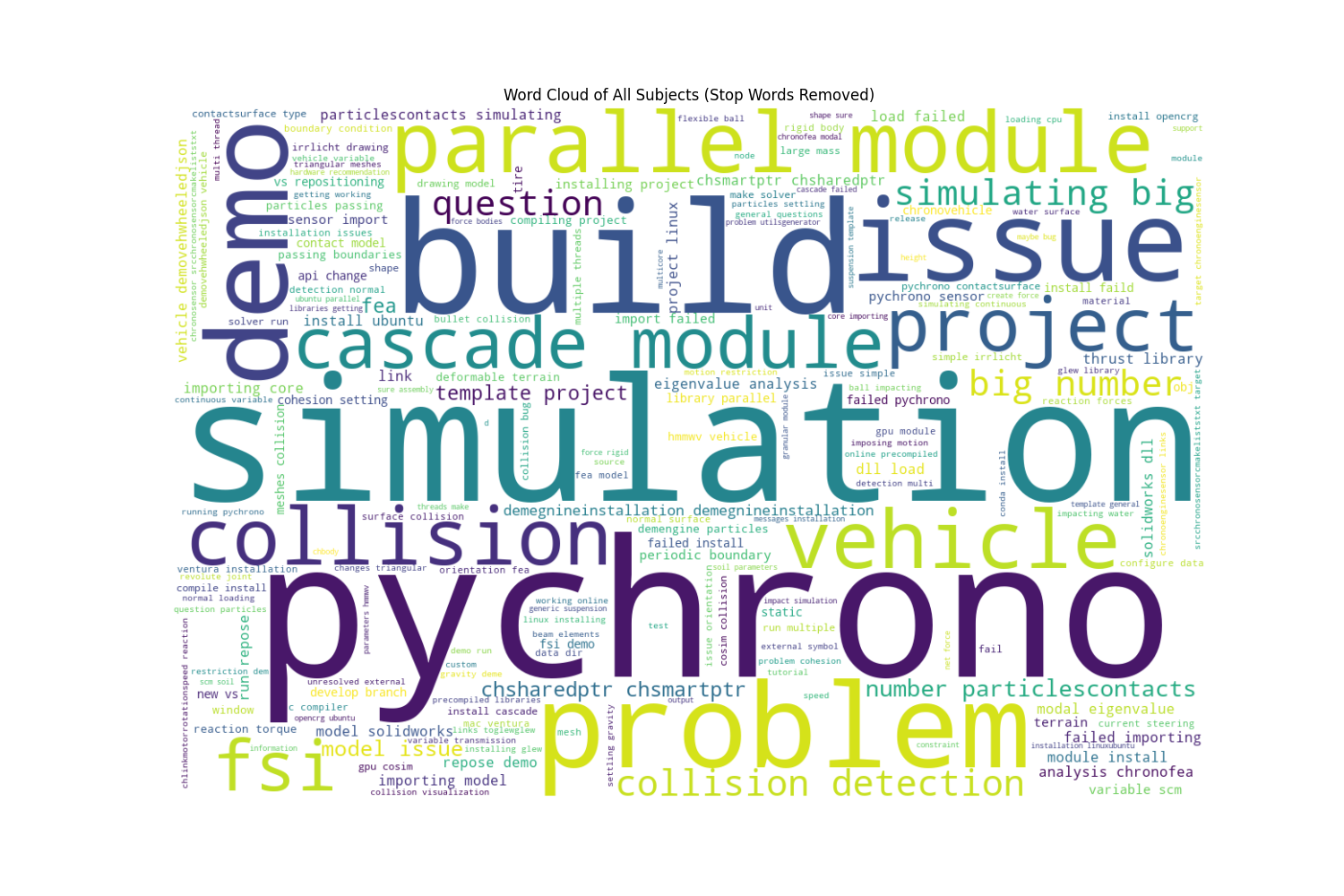}
    \caption{The word cloud of Project Chrono Forum}
    \label{fig:word_cloud}
\end{figure}

As detailed in Section \ref{sec:chrono}, PyChrono is extensively utilized by our users and serves as a multi-physics simulator catering to a wide range of applications. Given its extensive use, the structure of Project Chrono has evolved to accommodate complex simulations, making it inherently intricate. As of May 2024, Project Chrono has released version 9.0.0, continually improving its capabilities. To systematically identify the challenges faced by users working with PyChrono, we conducted an in-depth analysis of the discussions in the Project Chrono Google forum. Up to May 2024, there have been a total of 2,576 multi-round conversations in the forum. Figure \ref{fig:word_cloud} shows a word cloud representing the most frequently discussed key words. Using Latent Dirichlet Allocation (LDA) \cite{sievert2014lda} to analyze the subject lines of these conversations, we identified three primary topics:

\begin{enumerate}
    \item Case-by-Case Simulation Setup Challenges: The largest segment, accounting for $38\%$ of the discussions, focuses on the specific challenges encountered when setting up simulations with PyChrono. As users employ Chrono to configure highly complex simulations, they frequently engage in discussions about achieving correct configurations, handling unexpected simulation behaviors, and enhancing performance for particular scenarios. The LDA analysis underlines that a significant number of these challenges involve collision detection, configuring the dynamics of physical interactions, and ensuring accurate motion and detection of simulated bodies. These aspects are critical due to the real-world complexity of the simulations that users aim to model.
    \item API Usage Problems: Representing about $37\%$ of the forum discussions, this category highlights the difficulties users face with the extensive array of APIs provided by PyChrono. Despite continuous enhancements and optimization by the developers, the complexity inherent in the physics simulations has led to a vast codebase, currently comprising 18,320 APIs (including functions, variables, typedefs, enumerations, and related functions) as per the latest version of Project Chrono \cite{chronoAPIWebSite}. This extensive API surface, while robust, poses significant usability and accessibility challenges, particularly for newcomers or those engaged in complex simulation tasks. Users often report struggles in identifying the appropriate functions or employing them effectively within their projects. Additionally, the LDA findings suggest that these issues are compounded when addressing bugs, especially in simulations involving terrain and vehicles. Commonly reported API-related challenges include difficulties with tire models, vehicle APIs, collision detection APIs, and boundary conditions, as well as issues arising from cross-operating system compatibility.
    \item Module Installation and Compilation Issues: Making up approximately $20\%$ of the conversations, this topic addresses the issues related to the installation and compilation of various PyChrono modules. Users regularly encounter obstacles due to incorrect installation procedures or compatibility problems, which can severely impede their progress in starting with PyChrono. The LDA analysis draws attention to the fact that these discussions frequently center on specific modules, notably those involving the PyChrono sensor module and CUDA dependencies. These technical hurdles are significant barriers for users attempting to utilize the full capabilities of PyChrono in their simulation endeavors.
 
\end{enumerate}

\paragraph{The limitations of current LLMs for Chrono-related problems}
Despite advancements in large language models (LLMs) like GPT-4, they struggle to effectively address Chrono-related problems, often exacerbating complexities rather than providing practical solutions. While some LLMs exhibit a baseline understanding of PyChrono, as shown in Table \ref{tab:my_label}, their capabilities remain inadequate for generating reliable simulation codes or handling complex tasks in PyChrono.

\begin{table}
    \centering
    \begin{tabular}{ccccccc}
        Model & GPT-4  & GPT3.5  & LLaMA3-8B & Phi3-mini  & Mistral-7B & CodeGemma-7B \\
         Chrono Introduction& yes  & yes & yes &yes  &yes &yes \\
         Chrono API&  yes  & yes  & no  & yes & yes & yes \\
         &  &  &  &  & &\\
    \end{tabular}
    \caption{Baseline Chrono knowledge in various LLMs.}
    \label{tab:my_label}
\end{table}

While these models demonstrate some familiarity with the Chrono framework, particularly its introduction and basic APIs, their ability to generate accurate PyChrono simulation codes or adapt to more recent API changes (Table \ref{tab:api_changes}) is notably limited. For instance, they often fail to account for updates in API naming conventions or methods for solver settings, visualization, and collision handling.

\begin{table}[h!]
\centering
\begin{tabular}{|l|l|}
\hline
\textbf{Old API} & \textbf{New API} \\ \hline

\multicolumn{2}{|c|}{\textbf{Basic Usage}} \\ \hline
\texttt{chrono.ChVectorD(...)} & \texttt{chrono.ChVector3d(...)} \\ \hline
\texttt{chrono.ChQuaternionD(...)} & \texttt{chrono.ChQuaterniond(...)} \\ \hline
\texttt{chrono.ChCoordsysD(...)} & \texttt{chrono.ChCoordsysd(...)} \\ \hline
\texttt{sys.Set\_G\_acc(...)} & \texttt{sys.SetGravitationalAcceleration(...)} \\ \hline
\texttt{box1.SetPos\_dt(...)} & \texttt{box1.SetPosDt(...)} \\ \hline
\multicolumn{2}{|c|}{\textbf{Collision and Contact System}} \\ \hline
\texttt{chrono.ChCollisionSystemBullet(...)} & \texttt{sys.SetCollisionSystemType(...)} \\ \hline
\texttt{ground.SetBodyFixed(...)} & \texttt{ground.SetFixed(...)} \\ \hline
\texttt{ground.SetCollide(...)} & \texttt{ground.EnableCollision(...)} \\ \hline
\texttt{sys.GetNcontacts(...)} & \texttt{sys.GetNumContacts(...)} \\ \hline
\texttt{chrono.ChMaterialSurfaceNSC(...)} & \texttt{chrono.ChContactMaterialNSC(...)} \\ \hline
\texttt{chrono.ChMaterialSurfaceSMC(...)} & \texttt{chrono.ChContactMaterialSMC(...)} \\ \hline
\texttt{chrono.CastToChMaterialCompositeNSC(...)} & \texttt{chrono.CastToChContactMaterialCompositeNSC(...)} \\ \hline
\multicolumn{2}{|c|}{\textbf{Solver Settings}} \\ \hline
\texttt{sys.SetSolverMaxIterations(...)} & \texttt{sys.GetSolver().AsIterative().SetMaxIterations(...)} \\ \hline
\texttt{sys.SetSolverForceTolerance(...)} & \texttt{sys.GetSolver().AsIterative().SetTolerance(...)} \\ \hline
\multicolumn{2}{|c|}{\textbf{Visualization}} \\ \hline
\texttt{chrono.ChFrameD(...)}&\texttt{chrono.ChFramed(...)}\\ \hline
\texttt{pend\_2.SetFrame\_COG\_to\_REF(...)} & \texttt{pend\_2.SetFrameCOMToRef(...)} \\ \hline
\texttt{pend\_2.SetFrame\_REF\_to\_abs(chrono.ChFrameD(...))} & \texttt{pend\_2.SetFrameRefToAbs(chrono.ChFramed(...))} \\ \hline
\texttt{chrono.ChCylinderShape(...)} & \texttt{chrono.ChVisualShapeCylinder(...)} \\ \hline
\texttt{ground.AddVisualShape(...)} & \texttt{ground.AddVisualShape(..., chrono.ChFramed(...))} \\ \hline
\texttt{veh.ChIrrGuiDriver(...)} & \texttt{veh.ChInteractiveDriverIRR(...)} \\ \hline

\end{tabular}
\caption{Part of the API changes from old to new versions in PyChrono}
\label{tab:api_changes}
\end{table}

We observe that while some models are pre-trained on general PyChrono documentation, their performance on Chrono-specific tasks, particularly the generation of correct and efficient simulation codes, is insufficient (see Section \ref{sec:pretrain}). Addressing these limitations requires more than prompt engineering or learning techniques. Considering the complexity of PyChrono and the need for accurate, domain-specific outputs, we opted to fine-tune existing LLMs. Fine-tuning enables a more targeted enhancement of the model's capabilities, ensuring it aligns closely with PyChrono’s requirements and significantly improves user experience for Chrono-related tasks.

%% file: sections/Methodology.tex
\subsection{Problem Statement}
\begin{figure}
    \centering
    \includegraphics[width=15cm]{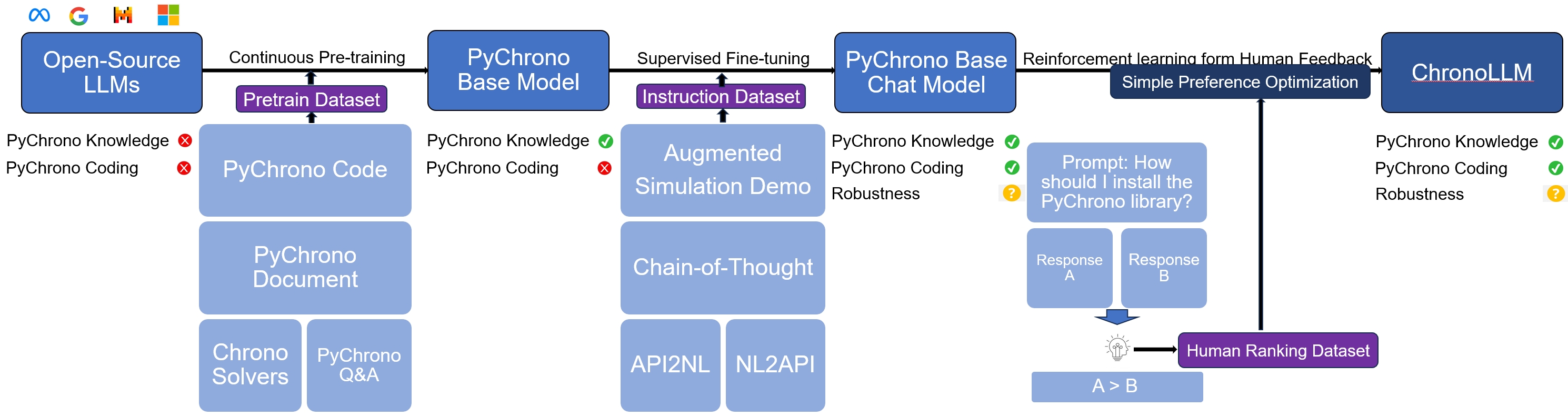}
    \caption{The whole pipeline of ChronoLlama to customize open-source LLMs for PyChrono tasks.}
    \label{fig:pipline}
\end{figure}

The whole pipeline of ChronoLlama is shown in Fig \ref{fig:pipline}, in which the training stages generally include two steps: continued pre-train and instruction fine-tuning.

\subsection{Selection of Base Models}  

Current large language models (LLMs) can be broadly categorized into two structural types: sparse structures, commonly referred to as Mixture of Experts (MoE), and dense structures. Examples of MoE models include the Mixtral family \cite{jiang2024mixtral} by Mistral AI and GPT-4. Dense models include LLaMA3-70B, GPT-4-O, and Gemma2-27B.

MoE models activate only a subset of their parameters (experts) for each input, enabling significant reductions in computational costs during inference. This allows for scaling to extremely large models without a proportional increase in computational requirements. While MoE models often offer faster inference at comparable sizes, they tend to lag behind dense models in accuracy. Moreover, training and managing MoE models can be complex, potentially leading to inefficiencies and reduced performance consistency. Dense models, in contrast, are easier to train and provide more robust, uniform performance, albeit at a higher computational cost during inference.

Considering these trade-offs, and the specific requirements of PyChrono-related tasks—which include accurate code generation and strong knowledge retrieval for Chrono-specific APIs—we selected dense models. The chosen models were evaluated based on their performance in widely recognized benchmarks, such as the HumanEval benchmark for coding tasks \cite{chen2021evaluating1} and the MMLU benchmark for general reasoning tasks \cite{hendrycks2020mmlu}.

In this study, we adopt state-of-the-art dense LLMs from leading AI research groups:
\begin{itemize}
    \item \textbf{GPT-4O} and its lightweight variant \textbf{GPT-4O mini} for their exceptional coding performance and general reasoning capabilities,
    \item \textbf{LLaMA3-70B}, the latest large-scale dense model offering strong performance on diverse tasks,
    \item \textbf{Gemma2-27B}, a high-performing model built on the latest advancements in the Gemma series \cite{google2024gemma}.
\end{itemize}

The decision to exclude some models, such as CodeLLaMA, stems from their limitations. Although CodeLLaMA is specialized for coding tasks, it is based on the older LLaMA2 architecture and, as of May 2024, underperforms compared to LLaMA3-70B on the HumanEval benchmark. Similarly, models like StarCoder \cite{li2023starcoder_LLM}, despite their multi-language coding specialization, exhibit insufficient performance for our use case.

The consistent accuracy, reliability, and robust performance of dense models make them the ideal choice for fine-tuning on PyChrono-related tasks. The simplicity of training dense models and their superior generalization across tasks further support their selection over MoE-based alternatives.

\subsection{Continual Pretraining}

The pretraining stage for large language models (LLMs) is predominantly unsupervised, enabling the models to process vast datasets and learn underlying patterns, structures, and knowledge. This process typically involves causal language modeling (CLM), where the model is trained to predict the next token in a sequence based on preceding tokens. During this stage, the model builds foundational knowledge that underpins its capabilities across a wide range of tasks.

In our approach, we continue pretraining the model using the standard CLM task. Given an input token sequence $\bm{x} = (x_0, x_1, x_2, \ldots)$, the model is trained to predict the next token $x_i$ in an autoregressive manner. The objective is to minimize the negative log-likelihood, defined as:

\begin{equation}\label{eq_clm}
    \mathcal{L}_{\textrm{CLM}} (\Theta) = \mathbb{E}_{\bm{x} \sim \mathcal{D}_{\textrm{PT}}}\left[ -\sum_i \log p(x_i \mid x_0, x_1, \ldots, x_{i-1}; \Theta)\right]
\end{equation}

Here, $\Theta$ represents the model parameters, $\mathcal{D}_{\textrm{PT}}$ is the pretraining dataset, $x_i$ is the token to be predicted, and $x_0, x_1, \ldots, x_{i-1}$ form the context.

It is widely assumed that LLMs acquire knowledge during pretraining, and certain emergent abilities—such as advanced comprehension, reasoning, and language generation—become apparent only when the pretraining loss falls below a critical threshold \cite{du2024understandemergent}. These emergent abilities significantly enhance the model’s capacity to follow instructions during fine-tuning. To ensure that the model gains specialized knowledge relevant to Chrono, we employ continual pretraining on Chrono-related materials.

Continual pretraining involves further training a pretrained model on new, domain-specific data. However, this approach presents challenges, particularly the risk of catastrophic forgetting (CF) \cite{french1999catastrophic}, where the model loses previously acquired knowledge while adapting to new data. To mitigate this risk, several strategies can be employed \cite{wu2022pretrained}:

\begin{itemize}
    \item \textbf{Warmup Strategies}: Gradually increasing the learning rate at the start of training helps the model adapt to new data without overwriting prior knowledge \cite{gupta2023continualwarm}.
    \item \textbf{Regularization Techniques}: Methods like elastic weight consolidation (EWC) penalize changes to critical weights, preserving previously learned information \cite{kirkpatrick2017overcoming}.
    \item \textbf{Replay Methods}: Mixing new data with previously seen data during training reinforces older knowledge while integrating new information \cite{lopez2017gradient}.
\end{itemize}

For our application, only the warmup strategy is feasible because, for most open-source LLMs, only the model weights are available, while the original training data is not. Additionally, regularization techniques can degrade performance on new tasks, making them unsuitable for our objectives. By employing the warmup strategy, we can continue pretraining LLMs to imbue them with specialized Chrono-related knowledge while ensuring the retention of their general capabilities. This approach ensures the model becomes more proficient in domain-specific tasks without compromising its broad utility.

\subsubsection{Continual Pretraining Dataset}

To effectively perform continual pretraining, it is essential first to assess whether it is necessary. This involves evaluating the model's current performance on PyChrono-related tasks. If the model demonstrates insufficient knowledge or poor performance, continual pretraining is warranted to enhance its capabilities.

Once the need for continual pretraining is established, the next step is to curate a suitable dataset. The following components are integral to creating a comprehensive pretraining dataset:

\begin{enumerate}
    \item \textbf{PyChrono Code Examples}:  
    This component involves cleaning and extracting representative code samples from various PyChrono projects using scripts. The dataset should cover a wide range of functionalities and use cases, ensuring diverse and high-quality learning material.

    \item \textbf{PyChrono Documentation}:  
    Include foundational documents, such as installation guides, introductory materials, and basic tutorials. To provide more depth, it is important to supplement these with advanced documentation covering complex concepts and APIs.

    \item \textbf{PyChrono Q\&A}:  
    Compile questions and answers from PyChrono user forums and support channels. This data helps the model understand common issues, troubleshooting methods, and practical applications. Although the current collection may be limited, efforts should be made to expand this dataset for better problem-solving capabilities.

    \item \textbf{Chrono Solvers}:  
    Although not immediately essential, materials related to Chrono solvers—such as solver documentation, examples, and user experiences—can be included to enhance the model's understanding of simulation and solver techniques. These materials will become increasingly valuable for complex simulation tasks.
\end{enumerate}

By integrating these components, the dataset ensures the model receives a comprehensive and in-depth education in PyChrono. Table~\ref{tab:pretrain_samples} provides examples from each category, formatted in JSON for compatibility with LLM training pipelines.

\begin{table}[h!]
\centering
\renewcommand{\arraystretch}{1.5} 
\begin{tabular}{|p{4cm}|p{10cm}|}
\hline
\textbf{Category} & \textbf{Sample} \\ \hline

\textbf{PyChrono Code Examples} & 
\texttt{
\{ "text": "import pychrono as chrono  import pychrono.fea as fea  import pychrono.pardisomkl as mkl..." \}
} \\ \hline

\textbf{PyChrono Documentation} & 
\texttt{
\{ "text": "Overview of vehicle modeling and simulation The Chrono::Vehicle module provides templates for various..." \}
} \\ \hline

\textbf{PyChrono Q\&A} & 
\texttt{
\{ "text": "Is Chrono free? The entire Chrono software infrastructure is open source and released under a BSD-3 license. ..." \}
} \\ \hline

\textbf{Chrono Solvers} & 
\texttt{
\{ "text": "// PROJECT CHRONO - http://projectchrono.org // Chrono solvers based on Eigen iterative linear solvers..." \}
} \\ \hline

\end{tabular}
\caption{Examples of JSON-formatted samples for continual pretraining in different categories.}
\label{tab:pretrain_samples}
\end{table}

This structured approach to dataset preparation will significantly enhance the model’s ability to generate accurate PyChrono code and provide reliable, domain-specific knowledge. As a result, the model's overall performance on PyChrono-specific tasks will improve, addressing both general and advanced use cases effectively.

\section{Fine-Tuning Methodology}

Fine-tuning is a crucial step in adapting large language models (LLMs) to specialized tasks, such as integrating the PyChrono library. Unlike pretraining, which focuses on learning general patterns from vast datasets, fine-tuning adjusts a pre-trained model's parameters to optimize its performance on specific datasets or tasks. This step is often the most critical for leveraging the capabilities of LLMs in domain-specific applications.

Chain-of-Thought (CoT) reasoning and its structured variants \cite{li2023structuredCOT} have proven particularly effective in enhancing LLM generation abilities. Fine-tuning builds upon these foundations, typically employing supervised learning to pair specific instructions with expected outputs. This process enables models to refine their reasoning and task execution capabilities.

\subsection{Supervised Fine-Tuning (SFT)}

Supervised Fine-Tuning (SFT) involves adjusting all of a model's parameters to optimize its performance. This approach can achieve excellent results for task-specific applications but requires substantial computational resources. The objective of SFT is to minimize the loss calculated on the output portion of the training sequence:

\begin{align}
    \mathcal{L}_{\textrm{SFT}}(\Theta) = \mathbb{E}_{\bm{x} \sim \mathcal{D}_{\textrm{SFT}}} \left[ -\sum_{i \in \textit{\{output\}}} \log p(x_i \mid x_0, x_1, \ldots, x_{i-1}; \Theta) \right]
\end{align}

Here, $\Theta$ denotes the model parameters, $\mathcal{D}_{\textrm{SFT}}$ represents the fine-tuning dataset, and $\bm{x} = (x_0, x_1, \ldots)$ refers to the tokenized input sequence.

Although SFT provides comprehensive parameter updates, the computational cost can be prohibitive, particularly for large-scale LLMs. In many cases, domain-specific data is relatively limited compared to general-purpose datasets, making full fine-tuning unnecessary. Parameter-efficient fine-tuning (PEFT) provides a more resource-efficient alternative.

\subsection{Parameter-Efficient Fine-Tuning (PEFT)}

PEFT methods refine a subset of the model's parameters or introduce lightweight adaptations to improve efficiency \cite{he2021towards}. These techniques have been widely adopted in LLM applications and include:

\begin{itemize}
    \item \textbf{Adapters}: Trainable modules inserted into the model's architecture \cite{hu2023adapter, zhang2023adapter2}.
    \item \textbf{Soft Prompts}: Task-specific embeddings added to the input sequence \cite{lester2021softprompt}.
    \item \textbf{Selective Updates}: Methods like BitFit \cite{zaken2022bitfit}, which update only specific parameters, such as bias terms, while freezing the rest of the model.
\end{itemize}

Among PEFT techniques, Low-Rank Adaptation (LoRa) \cite{hu2022lora} stands out as a particularly effective method.

\subsubsection{LoRa and Its Variants}

LoRa employs low-rank decomposition to parameterize weight updates. For a pre-trained weight matrix $W_0 \in \mathbb{R}^{d \times k}$, the weight update is defined as:

\begin{equation}
W_0 + \Delta W = W_0 + BA
\end{equation}

Here, $B \in \mathbb{R}^{d \times r}$ and $A \in \mathbb{R}^{r \times k}$, where $r \ll \min(d, k)$. During training, $W_0$ remains frozen, and only $A$ and $B$ are trainable. The modified forward pass for an input vector $x$ is:

\begin{equation}
h = W_0 x + \Delta W x = W_0 x + BA x
\label{eq:lora}
\end{equation}

LoRa is ideal for tasks requiring subtle adaptations to the attention mechanism. Variants like LoRa+ \cite{hayou2024lora+}, QLoRa \cite{dettmers2024qlora}, and GaLoRa \cite{zhao2024galore} extend its capabilities through differential learning rates, quantization, and global attention adjustments.

\subsubsection{Advantages of PEFT Methods}

PEFT techniques offer significant benefits:

\begin{itemize}
    \item \textbf{Storage Efficiency}: Adds minimal parameters, reducing storage requirements.
    \item \textbf{Memory Efficiency}: Requires less memory, enabling training on resource-constrained devices.
    \item \textbf{Computational Efficiency}: Low-rank updates reduce overhead, allowing faster training and inference.
\end{itemize}

\subsection{Dataset Preparation for Fine-Tuning}

Preparing a robust dataset is critical for fine-tuning. For the ChronoLlama project, the dataset must balance descriptive text and executable code to effectively demonstrate the functionalities of PyChrono. 

\paragraph{Data Collection and Generation}

Data is sourced from a combination of automated and manual methods:
\begin{itemize}
    \item \textbf{Descriptive Text}: Generated using LLMs like GPT-4 to provide detailed overviews and explanations. Additional data can be harvested using web crawlers to extract relevant documentation and forum discussions.
    \item \textbf{Executable Code Samples}: Collected from official PyChrono documentation and reliable open-source repositories. LLMs like ChatGPT can also generate tailored code examples, covering both basic and advanced use cases.
\end{itemize}

\paragraph{Dataset Components}

The dataset is structured into four JSON files:
\begin{itemize}
    \item \textbf{pychrono\_sft\_sim.json}: Instructions and corresponding code outputs for simulation tasks.
    \item \textbf{pychrono\_sft\_COT.json}: Chain-of-Thought examples with step-by-step reasoning.
    \item \textbf{pychrono\_sft\_NL2API.json}: Maps natural language instructions to API calls.
    \item \textbf{pychrono\_sft\_API2NL.json}: Maps API calls to descriptive natural language explanations.
\end{itemize}

\paragraph{Structure and Format}

The dataset is structured in JSON format for seamless compatibility with machine learning pipelines. Each entry contains three fields: \textit{"instruction," "input,"} and \textit{"output."} These fields encapsulate the task description, optional input context, and the expected response. Below are examples from different dataset categories:

\subparagraph{Simulation Generation (\texttt{pychrono\_sft\_robot\_sim.json})}

This dataset focuses on generating Python code for PyChrono simulations, covering physics settings, model initialization, and visualization.

\begin{quote}
\begin{verbatim}
[
  {
    "instruction": "Simulate the Curiosity rover traversing rigid terrain using PyChrono.",
    "input": "Rover model: Curiosity, Terrain: Rigid",
    "output": "import pychrono as chrono\nimport pychrono.robot as robot\n..."
  }
]
\end{verbatim}
\end{quote}

\subparagraph{Chain-of-Thought (\texttt{pychrono\_sft\_COT.json})}

This dataset includes step-by-step reasoning to enhance the model's ability to logically approach tasks.

\begin{quote}
\begin{verbatim}
[
  {
    "instruction": "Explain the steps to build a basic Curiosity rover model in a PyChrono.",
    "input": "",
    "output": "1. Import necessary libraries...\n2. Create a simulation system...\n..."
  }
]
\end{verbatim}
\end{quote}

\subparagraph{Natural Language to API Mapping (\texttt{pychrono\_sft\_NL2API.json})}

This dataset translates natural language instructions into PyChrono API calls.

\begin{quote}
\begin{verbatim}
[
  {
    "instruction": "Generate PyChrono code for a body with material properties and texture.",
    "input": "",
    "output": "ground_mat = chrono.ChContactMaterialNSC()\n..."
  }
]
\end{verbatim}
\end{quote}

\subparagraph{API to Natural Language Mapping (\texttt{pychrono\_sft\_API2NL.json})}

This dataset maps API calls to descriptive explanations, aiding comprehension.

\begin{quote}
\begin{verbatim}
[
  {
    "instruction": "Explain the following API: ground_mat = chrono.ChContactMaterialNSC()\n...",
    "input": "",
    "output": "Create a ground body with specified dimensions and material properties."
  }
]
\end{verbatim}
\end{quote}

\paragraph{Key Features}

The dataset is designed for:
\begin{itemize}
    \item \textbf{Compatibility}: JSON format ensures integration with standard pipelines.
    \item \textbf{Diversity}: Covers simulation generation, reasoning, and API understanding.
    \item \textbf{Scalability}: Can be expanded with new tasks or categories.
\end{itemize}

This concise, structured dataset ensures the model is well-equipped for PyChrono-related tasks.

%% file: sections/Numericaltest.tex
In this section, we will investigate the performance of the customized LLMs in PyChrono-related domain, here we only benchmark on the simulation digital-twin generation.

\subsection{Metrics for Evaluating Digital Twin Generation}

Evaluating Large Language Models (LLMs) presents significant challenges, particularly in the context of code generation tasks \cite{chang2024LLMevaluation, liu2024LLMcodeevaluation}. Generally, evaluation metrics can be categorized into three main types: similarity-based methods, execution-based methods, and LLM-as-judge approaches.

\textbf{Similarity-Based Methods.} These metrics assess the generated output by comparing it to reference code. A commonly used metric is \textbf{BLEU (Bilingual Evaluation Understudy)} \cite{papineni2002bleu1, EVTIKHIEV2023bleu2}, which measures the n-gram overlap between the generated and reference texts. \textbf{CodeBLEU} extends BLEU by incorporating syntax and semantic elements specific to code, providing a more nuanced evaluation for programming tasks. Additionally, the \textbf{ROUGE (Recall-Oriented Understudy for Gisting Evaluation)} family \cite{ganesan2018rouge} focuses on recall-based measures to evaluate the quality of summaries, and it has been adapted for assessing code generation by evaluating the overlap of essential components between the generated and reference code.

\textbf{Execution-Based Methods.} These metrics evaluate the functional correctness of the generated code by executing it against predefined tests. The metric \textbf{$\text{pass@}k$} \cite{chen2021evaluating1, xu2022evaluating2, liu2024evaluating3} represents the probability that at least one out of $k$ generated samples passes all unit tests, thereby indicating functional accuracy. Another important metric is \textbf{$\text{compile@}k$}, which checks whether the generated code successfully compiles, thereby ensuring syntactic correctness. These execution-based methods provide a direct measure of whether the generated code performs as intended.

\textbf{LLM-as-a-Judge Approach.} Beyond traditional metrics, the LLM-as-a-judge \cite{zheng2023LLMasJudge} paradigm leverages another LLM to evaluate the quality of the generated code. In our previous work \cite{wang2024simbench}, we introduced the \textbf{J-LLM Judge}, which utilizes reference code and API documentation to assess performance. Our findings indicate that the J-LLM Judge offers a more reliable evaluation metric compared to similarity-based methods like CodeBLEU and ROUGE-L\textsubscript{Sum}. Furthermore, it exhibits a stronger correlation with $\text{pass@}k$ than with $\text{compile@}k$, suggesting its effectiveness in evaluating both the functional and semantic quality of the generated code.

\textbf{Comprehensive Evaluation.} In the numerical evaluation section, we employ all the aforementioned metrics to thoroughly assess the performance of the LLM in generating digital twins. This multifaceted approach ensures a robust and comprehensive evaluation, capturing both the syntactic and functional aspects of the generated code.

\subsection{Test Models, Baselines, and Assumptions}

In this section, we introduce the baseline models employed for comparison in tasks related to PyChrono. Building upon the pretrained Large Language Models (LLMs) evaluated in our previous study \cite{wang2024simbench}, we select top-performing pretrained models and implement two distinct approaches to customize and embed PyChrono-related data. The first approach utilizes in-context learning, which leverages well-written API documentation and PyChrono-specific examples to enhance the model's understanding and generation capabilities without altering its underlying parameters. The second approach involves fine-tuning the models with PyChrono-specific data, allowing the models to adapt their weights based on the specialized dataset.

The selected models vary in scale and include both open-source and commercial options to ensure a comprehensive evaluation. Specifically, we perform fine-tuning on GPT-4o, GPT-4o-mini, Llama3-70B, Gemma2-27B. Additionally, we apply in-context learning to GPT-4o, GPT-4o-mini, and Llama3.1-70B. This selection encompasses a diverse range of model sizes and architectures, facilitating a robust comparison across different model capabilities and customization methods.

Evaluating large pretrained models introduces complexities distinct from those encountered with traditional deep learning models trained from scratch. In particular, it is challenging to categorically define whether the evaluation constitutes an in-distribution or out-of-distribution test. For example, widely recognized benchmarks such as MMLU and math reasoning tasks have been extensively optimized and can be considered "hacked" due to their widespread use and fine-tuning. In our context, Project Chrono is open-sourced and widely utilized, with all data publicly available on GitHub. The S-LLMs have already been pretrained on PyChrono-related data, which serves as the raw data for SimBench. Consequently, our fine-tuning and in-context learning datasets are synthesized from the same data sources, resulting in an inevitable overlap between the SimBench testing data and the training set. Therefore, testing on SimBench constitutes an \textbf{in-distribution test}, meaning that the models are evaluated on data that closely resembles their training data.

To ensure a fair and comprehensive comparison, we introduce an additional baseline referred to as the \textbf{"Hacked SimBench"}. This baseline involves extensively training the models on the exact same prompts and data used in SimBench. By doing so, we establish a performance benchmark for models that have been specifically trained on the precise data and prompts of SimBench. Comparing this baseline with other models and methods allows us to assess the relative effectiveness of our fine-tuning and in-context learning approaches under identical data conditions.

In summary, our evaluation framework encompasses a diverse set of models and customization methods, alongside carefully considered baselines, to rigorously assess the performance of LLMs in generating digital twins within the PyChrono ecosystem. This multifaceted approach ensures that our comparisons are both fair and indicative of the models' true capabilities in handling PyChrono-related tasks.

\begin{figure}[H]
    \centering
    \includegraphics[width=0.8\linewidth]{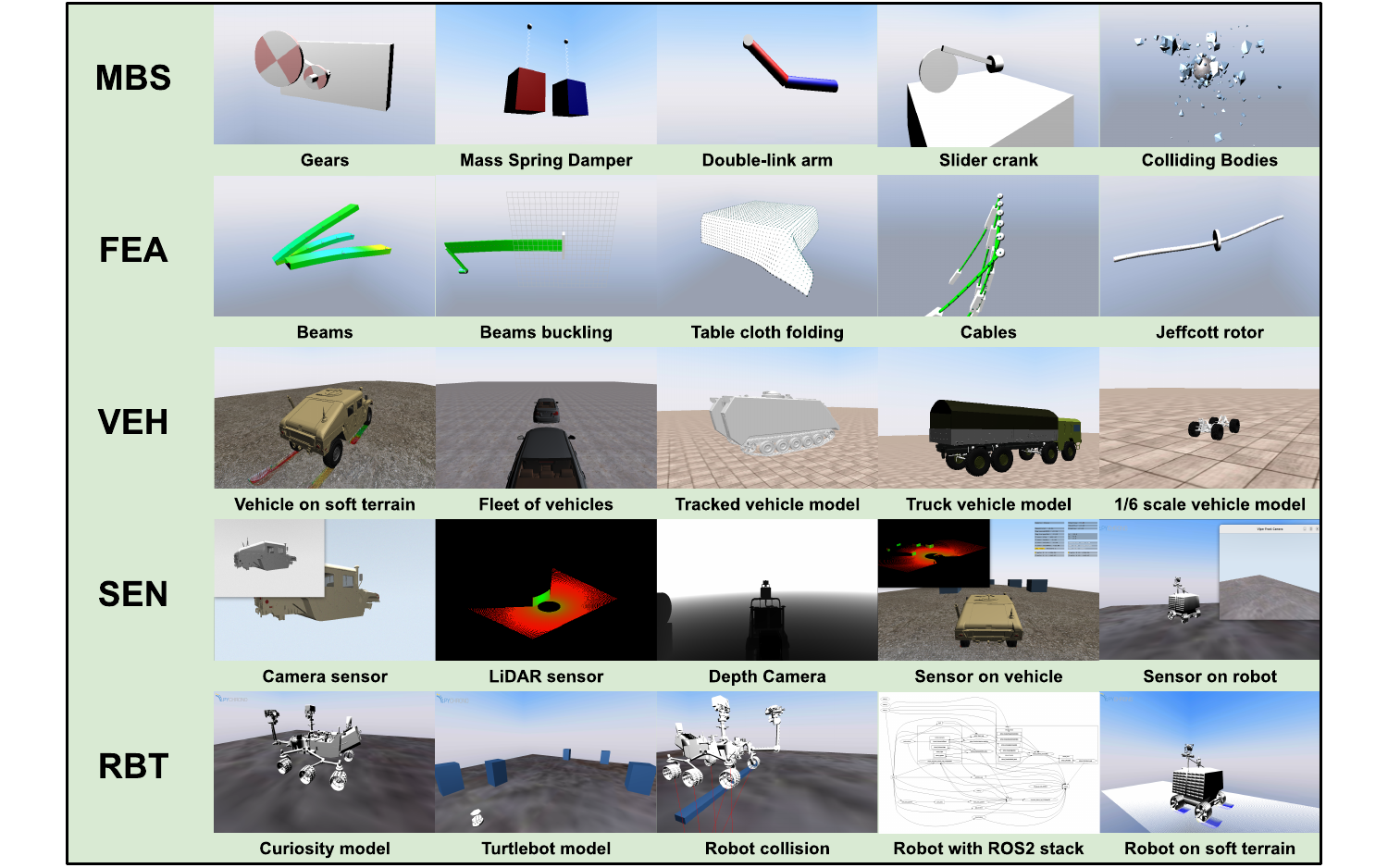}
    \caption{Overview of test environments categorized into various domains used in simulations. Each row represents a specific category, highlighting different scenarios for testing and evaluation of simulation models.}
    \label{fig:test_environment}
\end{figure}

\begin{figure}[H]
    \centering
    \includegraphics[width=0.9\linewidth]{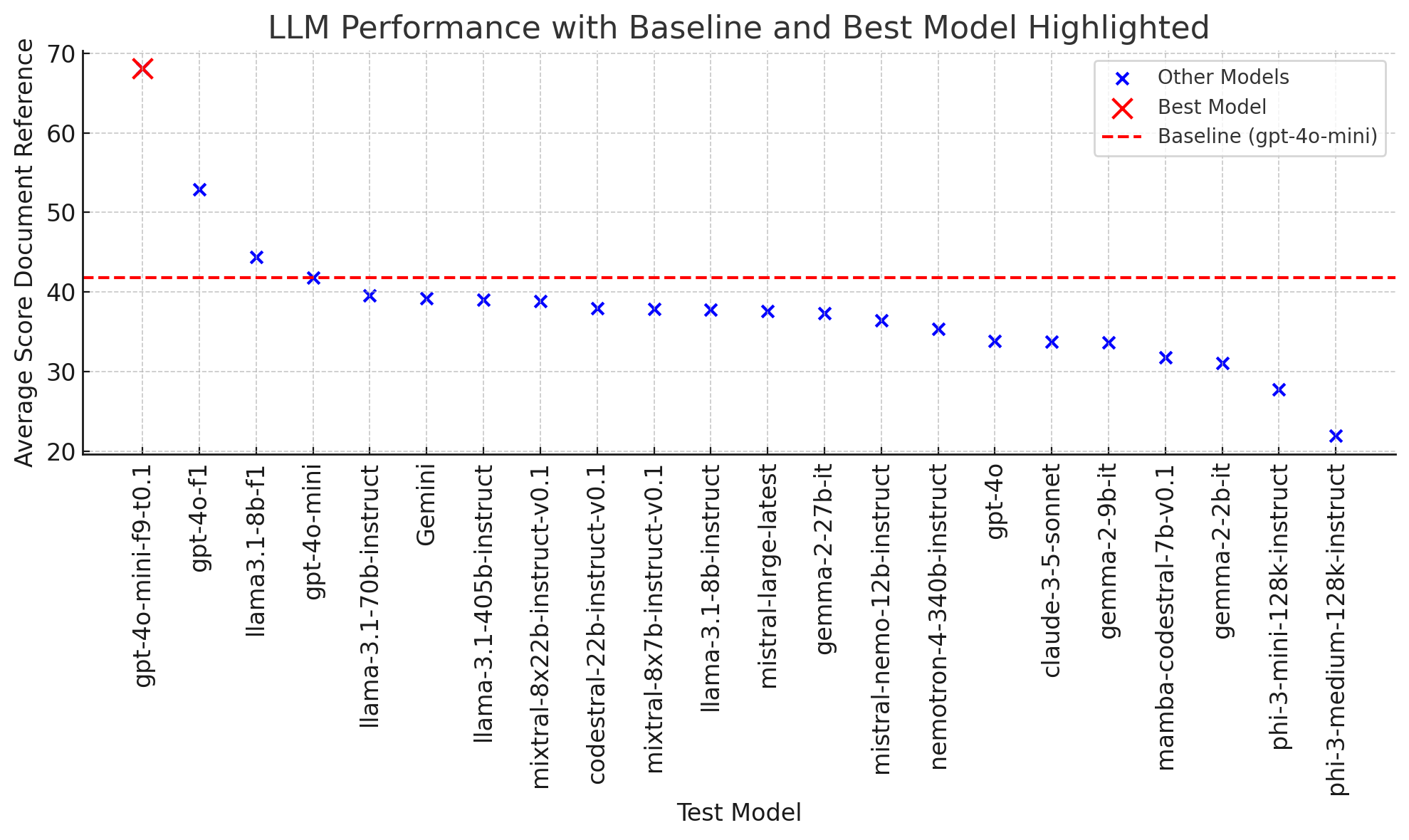}
    \caption{Comparison of LLM performance across different models based on the average document reference score. The fine-tuned model (\textit{gpt-40-mini-f9-t0.1}, marked in red) significantly outperforms all other models, achieving an average score close to 70. The baseline model (\textit{gpt-40-mini}, indicated by the red dashed line) achieves an average score of 40, showing a substantial improvement after fine-tuning. Other models are marked in blue for reference. This demonstrates the effectiveness of fine-tuning in enhancing the model's performance.}
    \label{fig:llm_performance_baseline_best}
\end{figure}

\begin{figure}[H]
    \centering
    \includegraphics[width=0.9\linewidth]{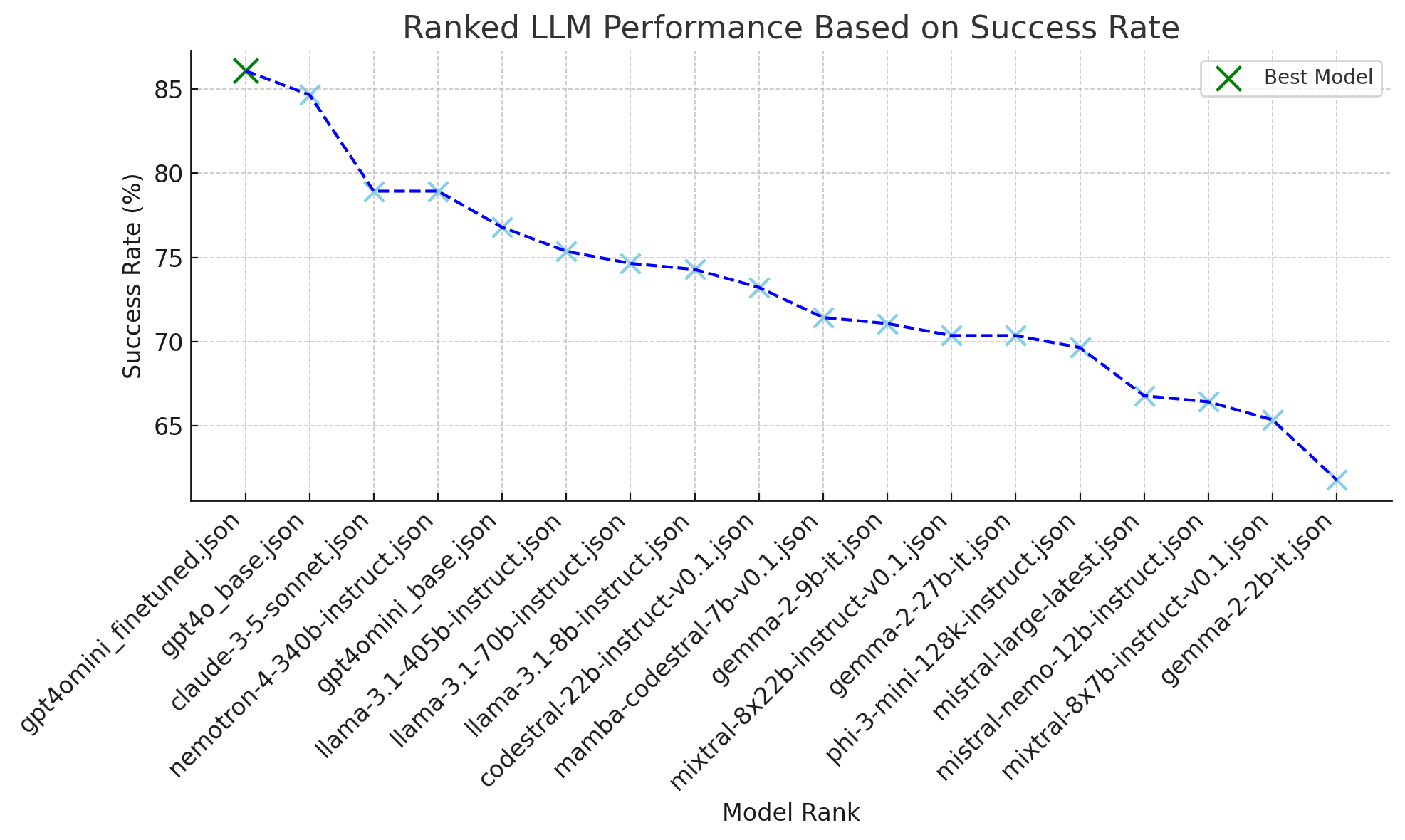}
    \caption{Ranked performance of various LLMs based on success rate. The fine-tuned model (\textit{gpt40mini\_finetuned.json}, marked with a green cross) achieves the highest success rate of approximately 85\%, far surpassing all other models, including the baseline (\textit{gpt40\_base.json}). The steady decline in success rate for other models, as indicated by the blue dashed line, highlights the fine-tuned model's superior robustness and accuracy.}
    \label{fig:ranked_llm_success_rate}
\end{figure}

%% file: sections/Conclusion.tex
This study evaluated the performance of customized Large Language Models (LLMs) in generating digital twins for PyChrono-related tasks, focusing on simulation-based digital twin generation. Fine-tuned models, such as \textit{gpt-40-mini-f9-t0.1}, significantly outperformed baseline and in-context learning approaches, achieving higher scores in metrics like BLEU, CodeBLEU, and pass@k. These results demonstrate the effectiveness of fine-tuning in enhancing functional and semantic accuracy.

To ensure fair evaluation, we introduced the "Hacked SimBench" baseline, which highlighted the fine-tuned models' superiority under identical data conditions. Additionally, the use of diverse test environments validated the models' ability to generalize across various PyChrono tasks.

In summary, fine-tuned LLMs are highly effective for digital twin generation in PyChrono domains, setting a strong foundation for further advancements in domain-specific LLM customization.

%% file: sections/Limitation.tex
While this project has successfully enhanced the PyChrono understanding and generation capabilities of the LLM models, several limitations must be acknowledged:
\begin{itemize}[htbp]
    \item Harmful and unpredictable content: Though our model can reject unethical queries, these models may still generate harmful or misaligned with human preferences and values. This issue may arise from biases in the training data or the models' inability to discern appropriate outputs in certain contexts.
    \item Insufficient data and training: Due to constraints in computing power and data availability, the training of the models may not be sufficient for optimal performance. It should be mentioned that the creation of simulation models also means that a mechanical model is underlying the simulation model, which is why the term (mechanical) model
is used in the following as a synonym for both kinds of simulation and mechanical models. To produce a (mechanical) model from natural language, an LLM needs more than just
general NLP capabilities. It also requires foundational knowledge in mechanical engineering, including geometry, kinematics, statics, and dynamics. Training an LLM solely on the
documentation of a simulation code would not be sufficient. As a result, there is still room for improvement in the pychrono understanding and code generation of the models.
    \item Lack of robustness: The models may exhibit brittleness in some situations, producing inconsistent or nonsensical outputs when faced with adversarial inputs or rare language phenomena.
    \item Comprehensive evaluation: Evaluating large language models is an important topic in the current era. While we have seen many evaluation benchmarks for LLMs, their comprehensiveness and appropriateness for LLMs should be well-studied and examined. A more diverse and comprehensive LLM evaluation dataset and benchmark will have a great positive effect on shaping the future of LLM research.
    \item Scalability and efficiency: Although we applied LoRA and quantization to make the model more accessible to a broader community, when combined with the original LLaMA, the models' large size and complexity can lead to difficulties in deployment, especially for users with limited computational resources. This issue may hinder the accessibility and widespread adoption of the models in various applications.
\end{itemize}

%% file: sections/Future_work.tex
In this work, the following directions will be investigated for future improvements:

\begin{enumerate}
    \item Unlearning: Although we know that the base model we used for fine-tuning contains old and wrong data information, we didn't directly deal with them but tried to overwrite it with the knowledge of the latest version of Project Chrono. It's still possible for the old wrong information to appear in the future inference of LLM. A possible way is called 'unlearning' \cite{yao2024LLMunlearning}, which is to let LLMs forget the wrong, improper information. We will try to unlearn the old data and then finetuned on the correct and new data.
    \item Multi-Modal LLMs: Recognizing the growing importance of digital twins in modern engineering, we will also develop multi-modal LLMs capable of processing images and videos to construct precise digital twins. These models will integrate visual data processing with textual and numerical information to create highly accurate and dynamic representations of physical systems. This capability will be pivotal for applications requiring real-time updates based on visual inputs, such as adaptive manufacturing processes, responsive urban planning, and personalized healthcare simulations.
    \item Enhanced Tool Interaction:We aim to significantly advance the integration capabilities of LLMs, enabling them to interact seamlessly with compilers and leverage external numerical computing resources. This initiative will focus on developing sophisticated interfaces that allow LLMs to dynamically interact with a variety of software tools, enhancing their utility and efficiency in real-world applications. This enhanced interaction promises to streamline the development process, reduce error rates, and accelerate the transition from code generation to deployment.
    \item Multi-Level Agent LLMs: Building on the success of multi-agent systems in robotics, we plan to design multi-level agent LLMs specifically tailored for the construction of simulation code. These systems will employ hierarchical decision-making structures, where agents at different levels manage specific aspects of the simulation framework, from low-level numerical computations to high-level scenario planning. This approach will facilitate more complex, adaptive, and robust simulation environments, mirroring the collaborative dynamics found in intelligent autonomous systems.
\end{enumerate}

%% file: sections/Appendix.tex
\begin{figure}[H]
    \centering
    \includegraphics[width=0.9\linewidth]{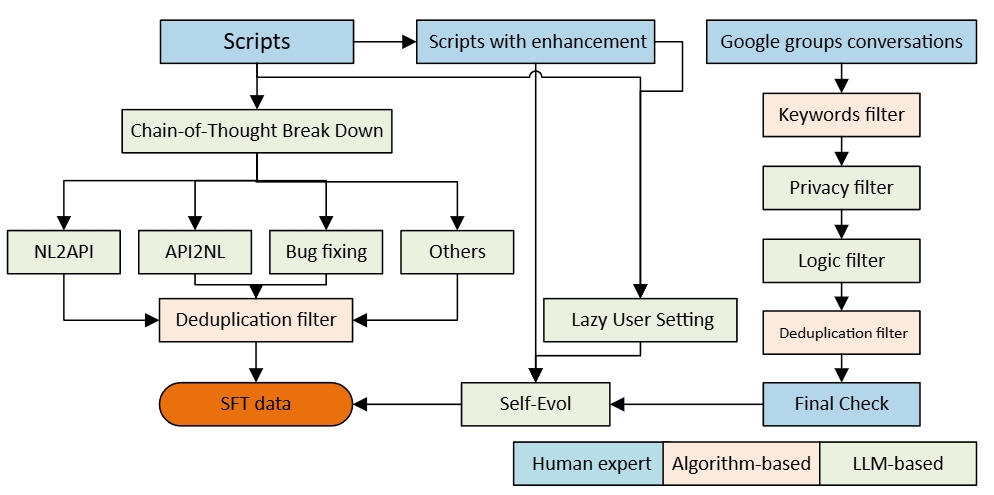}
    \caption{A comprehensive pipeline for synthesizing the Supervised Fine-Tuning (SFT) dataset. The process begins with the ingestion of scripts and Google Group conversations. Scripts are enhanced through various transformations such as Chain-of-Thought (CoT) breakdown, natural language to API (NL2API) conversion, API to natural language (API2NL) translation, bug fixing, and other enhancements. Google Group conversations are filtered through keyword extraction, privacy checks, and logic validation. Both data streams undergo deduplication and are checked for quality using Lazy User Setting and Self-Evolution mechanisms. Finally, the resulting dataset undergoes a final check before forming the SFT data, leveraging human experts, algorithm-based processes, and LLM-based techniques.}
    \label{fig:syn_pipeline}
\end{figure}

\begin{figure*}[ht]
	\centering
	\ovalbox{
		\begin{minipage}{1.0\linewidth}
			{\tiny
				\begin{lstlisting}[breaklines=true, numbers=none]
 You are a PyChrono expert tasked with evaluating a simulation script by comparing it against a reference script generated by experts. Your evaluation should consider both the accuracy of the script compared to the reference and adherence to best practices as outlined in the PyChrono API documentation.
					
 Here is the PyChrono code you need to evaluate:
 [The Start of Assistant's Answer]
 {code}
 [The End of Assistant's Answer]
					
 Here is the expert-generated reference code:
 [The Start of Reference Answer]
 {reference_code}
 [The End of Reference Answer]
					
 Use the following evaluation criteria and guidelines for point deduction:
  1. **Completeness (40 points total)**
   - Compare the provided code to the reference script. Deduct **15 points** for each missing essential component (e.g., system initialization, body creation, visualization) that is present in the reference script.
   - Deduct **10 points** if a component is present but lacks important details or is incorrectly configured compared to the reference.
   - Deduct **5 points** for minor omissions or slight deviations from the reference script.
  2. **Correctness (30 points total)**
   - Compare the code to the reference script. Deduct **15 points** for each incorrect use of a PyChrono API call that could lead to a change in simulation behavior.
   - Deduct **10 points** for logical errors in the code, such as incorrect joint initialization or wrong setting of body properties, especially if the reference script does it correctly.
   - Deduct **5 points** for minor inaccuracies or unnecessary API calls that deviate from the reference script.
  3. **Code Quality (10 points total)**
   - Evaluate the readability, structure, and documentation of the code against the reference script. Deduct **5 to 10 points** for poor readability, structure, or lack of meaningful variable names and formatting.
   - Deduct **5 points** for insufficient comments or failure to follow documentation best practices, especially if the reference script provides better documentation.
  4. **Efficiency (10 points total)**
   - Evaluate the efficiency of the code compared to the reference script. Deduct **5 points** for each instance of unnecessary calculations, redundant code, or inefficient use of APIs that is optimized in the reference script.
   - Deduct **3 points** for missing obvious optimization opportunities that the reference script implements.
  5. **Error Handling and Robustness (5 points total)**
   - Assess the error handling and robustness of the code. Deduct **5 points** for lack of basic error handling or failure to account for common issues that the reference script handles.
   - Deduct **3 points** for inadequate handling of edge cases compared to the reference script. 
  6. **Use of Visualization Tools (5 points total)**
   - Compare the use of visualization tools in the provided code to the reference script. Deduct **3 to 5 points** for incorrect or inadequate visualization setup as per the reference script.
   - Deduct **2 points** for minor visualization issues, such as suboptimal lighting or incomplete setup of visual elements, compared to the reference.
					
 Avoid position biases and ensure that the order in which the responses are presented does not influence your decision. Do not allow the length of the responses to influence your evaluation. Do not favor certain names of the assistants. Be as objective as possible.
					
 After providing your explanation, output the final score using the format '[[x]]' where x is the score assigned to the assistant's answer.
					
 Reference the PyChrono API documentation provided here: {api_documentation_link}
					
 Provide the evaluated score and a brief explanation of the deductions below:
				\end{lstlisting}
			}
		\end{minipage}
	}
	\caption{Instructions for using J-LLM with expert written API documentation and ground truth code.}
	\label{fig:eval_framework}
\end{figure*}

\begin{figure*}[ht]
	\centering
	\ovalbox{
		\begin{minipage}{1.0\linewidth}
			{\tiny
				\begin{lstlisting}[breaklines=true, numbers=none]
Your task is to generate high-quality, context-rich question-and-answer pairs based on the provided PyChrono code.
				
Imagine you are a **normal PyChrono user** encountering code from another language model. To resolve your issue, you must provide very concrete context when asking for help from another LLM.
				
Requirements:
- Generate {num_pairs} Q&A pairs that are clear, specific, and focused on key points from the context.
- Each question must be understandable without prior knowledge, including necessary background details from the context.
- Avoid vague references like "in this simulation" or "in the code", provide full context in the question.
				
Instructions:
1. **Question Generation**: Write a concise, clear question tied to a specific aspect of the code, ensuring enough context is given without revealing the answer. Imagine you're explaining the problem as a typical PyChrono user, needing to convey the full problem to get help from another LLM.
2. **Answer Generation**: Provide a precise answer based only on the provided context. Include explanations, corrections (if applicable), and detailed reasoning where necessary.
				
Output Format:
Generate the output in JSON format as follows:
{{
	"instruction": "<clear, context-rich question from the perspective of a *normal PyChrono user*, with full problem context>",
	"input": "",
	"output": "<detailed answer with explanations and code corrections (if needed)>"
}}
				
Example:
{{
	"instruction": "How can you set the collision system type in PyChrono to use the Bullet physics engine?",
	"input": "",
	"output": "To set the collision system type in PyChrono to Bullet, use `GetSystem()` to access the system and `SetCollisionSystemType()` to set it: `vehicle.GetSystem().SetCollisionSystemType(chrono.ChCollisionSystem.Type_BULLET)`."
}}
				
Ensure all questions are **specific** and **contextually rich**, with **no vague references** like "in the code". Provide clear, detailed answers that fully explain the issue and solution.
				
Context:
{markdown_content}
				\end{lstlisting}
			}
		\end{minipage}
	}
	\caption{Instructions for using LLM to synitize data.}
	\label{fig:syn1}
\end{figure*}

\begin{figure*}[ht]
	\centering
	\ovalbox{
		\begin{minipage}{1.0\linewidth}
			{\tiny
				\begin{lstlisting}[breaklines=true, numbers=none]
You're an expert in the PyChrono simulator. Your task is to generate question-and-answer pairs for the given context using the format below. 
When generating the question, imagine you are a user of the PyChrono simulation but do not have access to the given context. So DO NOT use phrases like 'in the script' or 'in the simulation' without given any context!!! 
Instead, ask about the problem with clear and detailed context directly!
For generating the answer, base it on the given text and be as detailed as possible.
					
Requirements:
- Create {num_pairs} question-and-answer pairs.
- Avoid phrases like 'in the script' or 'in the simulation' without given any context, be specific!
- The questions should be diverse, reflecting different ways users might phrase them. 
- Include best practices or common mistakes in the explanations where appropriate.
					
Format:
{{
	"instruction": "<diverse question, may include code>",
	"input": "<use input only when necessary>",
	"output": "<detailed explanation, may include code>"
}}
					
Markdown Content:
{markdown_content}
				\end{lstlisting}
			}
		\end{minipage}
	}
	\caption{Instructions for using LLM to synitize data.}
	\label{fig:syn2}
\end{figure*}

\begin{figure*}[ht]
	\centering
	\ovalbox{
		\begin{minipage}{1.0\linewidth}
			{\tiny
				\begin{lstlisting}[breaklines=true, numbers=none]
Your task is to generate question-and-answer pairs for a given PyChrono markdown file, focusing on **debugging tasks**. Follow the detailed instructions below for creating these pairs.
					
Requirements:
- Generate {num_pairs} pairs of questions related to **incorrect code** and provide detailed answers that describe the issues and how to fix them.
- Each question should focus on debugging or fixing errors in the given PyChrono code.
- The incorrect code must appear in the **instruction** section, framed as if a normal **PyChrono user** is asking for help with a bug or issue.
- The answer must describe what the error is and provide the corrected version of the code, along with a natural language explanation.
					
Rules for Bug Code Generation:
					
1. **API Misuse**: Common mistakes involve calling methods in the wrong order or using an API in a context where it's not valid.
	- **Example Error**: Calling `system.SetStep(0.01)` before the system is initialized.
					
2. **Misspelled API Names**: Simple typos or incorrect capitalization that lead to errors.
	- **Example Error**: Typing `ChSystemNSC` as `ChSystemNCS`; using 'chrono.ChVector3d' as 'chrono.ChVector3D'; using 'chrono.ChFramed(...)' as 'chrono.ChFrame(...)'.
					
3. **Wrong Parameter Types**: Passing values that are not valid for specific function parameters.
	- **Example Error**: Passing a string (`"1.0"`) instead of a float (`1.0`) to `SetMass()`.
					
4. **Incorrect Initialization**: Leaving out necessary steps like setting the mass or inertia of an object, leading to runtime issues.
	- **Example Error**: Failing to set the position of a `ChBody()`.
					
5. **Logic Errors**: Errors that don't crash the code but lead to unrealistic simulation behavior.
	- **Example Error**: Setting all joints as fixed, making the system immovable.
					
6. **Wrong Data Types or Values**: Using nonsensical values that make the simulation physically incorrect.
	- **Example Error**: Setting mass to a negative number.
					
7. **Unreasonable Time Step**: Setting a time step so large that it causes instability in simulations.
	- **Example Error**: `system.SetStep(1.0)` instead of a reasonable value like `0.01`.
					
Instructions:
1. Imagine you are a **normal PyChrono user** who has encountered a bug or error in the code and is seeking help **without knowing the exact issue**.
Again, the user should not know the exact error but should provide the code snippet that is causing the problem, framed as a question!!!
					
2. The incorrect code should appear in the **instruction** section of the output, and the **input** section should remain empty.
3. Provide a detailed natural language explanation of what the error is, why it occurs, and how to fix it, including the corrected version of the code.
4. Generate a diverse set of questions focusing on different types of errors and issues in the PyChrono code.
5. The code provided in the markdown file is ***absolutely correct***, and it should be used as the basis for generating the Q&A pairs.
					
Output Format:
Generate the output in JSON format as follows:
					
{{
	"instruction": "<incorrect code, framed as a user question>",
	"input": "",
	"output": "<detailed explanation of the error and the corrected code>"
}}
Example:
{{
	"instruction": "This code isn't working correctly, and I can't figure out why: ```python\n# Create Chrono system\nsystem = chrono.ChSystemNCS()\nsystem.SetCollisionSystemType(chrono.ChCollisionSystem.Type_BULLET)\nsystem.SetGravitationalAcceleration(chrono.ChVector3d(0, 0, -9.81))\n```",
	"input": "",
	"output": "The error here is a typo in the system initialization function. The correct function is `ChSystemNSC()` (Non-Smooth Contacts), not `ChSystemNCS()`. Here's the corrected code:\n\n```python\nsystem = chrono.ChSystemNSC()\nsystem.SetCollisionSystemType(chrono.ChCollisionSystem.Type_BULLET)\nsystem.SetGravitationalAcceleration(chrono.ChVector3d(0, 0, -9.81))\n```\nThe corrected code now initializes the system properly for non-smooth contacts."
}}
PyChrono Script Markdown file Content:
{markdown_content}
				\end{lstlisting}
			}
		\end{minipage}
	}
	\caption{Instructions for using LLM to synitize SFT data.}
	\label{fig:syn3}
\end{figure*}